# Global Software Engineering in the Age of GitHub and Zoom

James Herbsleb
Institute for Software Research
School of Computer Science
Carnegie Mellon University
herbsleb@cmu.edu

## Abstract
Much has changed since the inaugural ICGSE conference in 2006. Tools have improved, awareness of cultural differences is widespread, and developments such as the foregrounding of open source have all enhanced our ability to work across geographic divides. But the pervasive and profound impact of software in the world – especially for societal scale systems such as social media – forces new and deeply challenging responsibilities on both developers and academics. We must find better ways of incorporating ethics into our development practices and pay far more attention to harmful unintended consequences as deployed systems interact with and often disrupt crucial social systems.

## Introduction

I am grateful for the opportunity to reflect on how global software development (GSD) has changed in the 14 years since I gave the keynote at the first ICGSE conference in 2006 in Florianopolis. Back then, we were primarily concerned with such things as

- Understanding why a routine project became so difficult to manage and so prone to failure if the work was spread across different sites,
- Understanding how fundamentals of GSD such as cultural differences, time zone differences, outsourcing models impacted the work, and
- Understanding how "virtual" teams differ from collocated teams, and how disturbed projects should be managed

I would not regard any of these issues as definitively solved, but we have certainly made progress in all these, and many other, areas. I think many readers are familiar with much of this work, so I will not take up space recounting it, but rather, **my goals in this paper are to briefly summarize my perceptions of where we are now, to characterize how the changing face of software development generally has impacted GSD, and to look forward to new issues and challenges we must overcome.**

Because of the pervasive role software has come to occupy in the contemporary world, I find myself compelled to look beyond the narrow confines of software development to our broader impact and responsibility to society more generally. I think we can no longer be just technologists, even professionally responsible ones. We are uniquely positioned, as professionals and as academics, to insist that ethical considerations, not just profit motives, guide our path.

## What has changed since 2006?

There have been several important changes that make distributed development easier and more efficient. One important difference is that our tools are now much better. Git greatly simplifies parallel work and moving changes across different workspaces. GitHub has added a number of social and awareness features that create a "social coding" transparent environment that provides fodder for a great many social and technical inferences about the state of a project and the activities of other developers [2] [4] [5]. Badges, which have become quite popular, can effectively encourage good development practices [16]. The pervasiveness of social coding also carries the benefits of de facto standardization, as when a developer we were interviewing said of pull requests, that they have created a uniform "language of contribution."

Zoom and other contemporary teleconferencing applications, while imperfect (we've all seen our colleagues suddenly freeze from time to time) are mostly quite good, and allow remote colleagues to converse, see each other's expressions, and share windows and screens. Teleconferencing existed in 2006, but it was clunky, expensive, and unreliable. Now, with just a laptop and an internet connection, it is easy, inexpensive, and reliable. That is a game changer. Various asynchronous applications, such as Slack, have grown in popularity and usability. In the early days of GSD, there were only a few chat applications, and they were seen mostly as distracting and as productivity-killers until research began to show otherwise [8]. We now understand the role of asynchronous communication tools much better, and make good use of many different media [15].

I think we have also learned a lot about how to work effectively in geographically separated conditions. People no longer seem quite as surprised when colleagues in other parts of the globe say things that might be odd, confusing, or even insulting if uttered by a local. It is expected that customs and norms differ, and we are much more likely these days to give each other the benefit of the doubt, rather than reacting with hostility and alienation. Teams and organizations are so frequently international that we have exposure to people from many different cultures and make adjustments almost without thinking about it.

The foregrounding of open source – from inconsequential oddity to a key resource – and the increased exposure to its development style, have also enhanced nearly everyone's ability to work with others at a distance. While not every open source community is a model of comity, people have learned to work together effectively in such environments. They have also learned to use, and often to contribute to, many libraries and frameworks in the open source world. This means that many novice developers have been exposed to key technologies critical to their work before they are even hired. It also means that many developers have much in common in their understanding of how to write code and how to behave in a community of developers, although many barriers remain [13], particularly for women and underrepresented minorities [3].

Despite the many improvements, we have accrued challenges as well. One substantial challenge is simply the tremendous and expanding demand for software development and maintenance. CPU speeds have vastly increased, memory has become orders of magnitude cheaper, and nearly everything with an "on" switch has software in it. Software systems are becoming much larger and more complex, and there are far more of them. As we hurtle toward a world of the Internet of Things, nearly every "thing" needs software. Privacy concerns and bandwidth considerations will move more and more computation to the edge.

This rapidly increasing demand for software is projected by the Bureau of Labor Statistics[1] to lead to an increase in positions for software developers of 22% in the next decade, vastly outstripping the average growth rate for all occupations (5%). This growth rate means we need to add a net of 316,000 software developers over the next ten years. Current demand is already leading to workforce shortfalls. Many analyses show that our current rate of producing software engineers will fall far short of our future needs. This, in turn, creates two consequent challenges. To produce anything like the number of developers we will need, we need to find ways to make the field encouraging and accessible to women and minorities who are currently dramatically underrepresented. Women comprise less than 19% of software developers, while less that 6% are Black or African American, and about 5% are Hispanic or Latino[2]. We can't meet the workforce demands of the future while half or more of the potential talent is sidelined. We also need to find new and better ways to turn workers not trained specifically in computer science to handle the programming demands in their domain. End user programming is an active area of research, of course, and many non-software-engineers currently write code (e.g., accountants creating giant spreadsheets, physicists writing scripts to analyze and plot data). We need to help bring key software engineering ideas into other fields, to help people organize their code, make it more general, robust, and understandable (see, e.g., [9]). We need much more interdisciplinary work in this area.

## Intended effects and side effects

We also need to pay much more attention to the impact of our work. It was once the case that interacting with a computer was a distinct and limited activity, e.g., editing some text, using a CAD system, sending an email. Today, almost everything we do involves interacting with computers. Looking just at mobile phone use, devices which would have been considered supercomputers a decade ago, there is very little we can't do with the right app. From shopping, finding a date, watching or making videos, and on and on, we live in and through our computing devices.

The upshot of this evolution is that in the past, the way software was written had mostly a limited impact on small slices of our life. Today, while we have not yet been sealed away in a Matrix-like invented world, software shapes the possibilities and limitations of individuals, and profoundly influences how we function as a society. The decisions of tech companies, teams, and individual developers shape our experience of the world. While we generally don't notice it, as the fish (so goes the cliché) doesn't notice water, we are immersed in computing technology, and its functionality is primarily given by its software.

---

[1] https://www.bls.gov/ooh/computer-and-information-technology/software-developers.htm
[2] https://www.bls.gov/cps/cpsaat11.htm

*Developers create the world we all inhabit.* Roughly speaking, that world consists of intended effects and – *perhaps more importantly* – side effects of the software that is introduced. Let me illustrate with an example.

In the world before social media, there was a small number of primary media outlets. Since there were just a few "pipes" for information to be fed to the public, access was fairly carefully controlled, largely favoring the most accurate and thoughtful content, a process sometimes referred to as "social epistemology" [Rauch, 2018] We come to know about the world through an inherently social process, which can be thought of as a sort of funnel (see Figure 1).

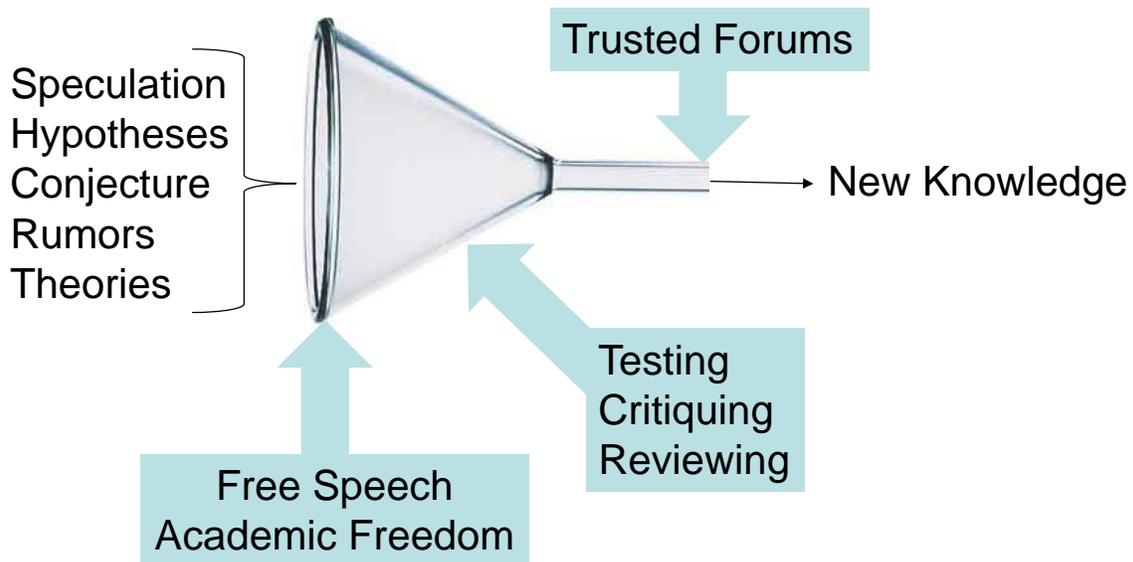

*Figure 1. A funnel model of social epistemology.*

Greatly oversimplifying, one could say the idea of social epistemology explains how we find or construct true knowledge of the world. Unfiltered speech, where creativity, speculation, and imagination are highly valued, must survive the funnel to be considered "knowledge." In the funnel, this free and unfiltered speech is subjected to criticism, vetted by knowledgeable reviewers, evaluated against data, and if it survives this process, enters a trusted forum where it is provisionally accepted as new knowledge. Academics are familiar with one version of the funnel, which is the review process for our peer-reviewed conferences and journals. It is an imperfect and sometimes frustrating process, but necessary if we are to have confidence in results. Many fields have versions of this funnel. Newspapers have editors, and rules such as "every report of a fact must be backed by two independent sources." All the elements that could be part of a story must pass through the funnel. Other funnels exist for nearly every type of knowledge work, e.g., intelligence communities, law enforcement, the legal system, medicine, management, etc.

Social media has fundamentally changed this picture, by substituting algorithms that select content in order to maximize engagement rather than ensure accuracy (see Figure 2). Facebook, Twitter, YouTube and all the rest have developed very sophisticated algorithms that use their all-encompassing data, along with every detail of nearly every user's actions on the platform, to show content most likely to keep everyone reading, viewing, posting, liking, commenting, interacting with the site, and hopefully clicking on targeted ads. The unfiltered content at the large end of the funnel is still freely generated, but now rather than undergoing a vetting process, it is filtered based only on whether it is likely to keep each user engaged. Each user and community receives content that holds its attention, with no particular concern for truth or rationality.

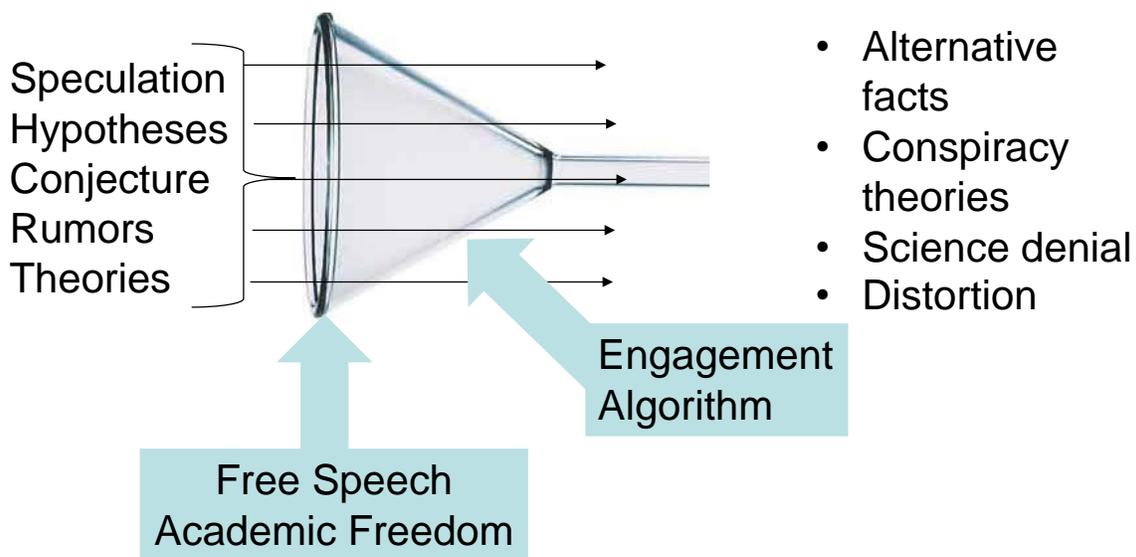

*Figure 2. Engagement algorithms and disintermediation have seriously damaged the social foundations of knowledge.*

Engagement algorithms have major side effects. Political content on YouTube, for example, tends to get more and more extreme as you follow the "up next" links [11]. Curiosity that leads a user to investigate the anti-vaxxer movement with Google search ends up going down a rabbit hole of one conspiracy theory after another. These sites make little more than token efforts to show you what is true and accurate. They have spent enormous resources, however, to show you what will keep you there, interacting with their site and generating ad revenue. The funnel of social epistemology gives way to the flood of visceral appeals, whatever it is that you want to hear and see and believe. As a consequence, false news travels faster and farther than the truth [17].

This is just one example of how the damaging impact of unintended consequences can overwhelm the benefits of the intended consequences. The intended consequence of engagement algorithms is to allow us to find more interesting content than we would otherwise find, and to allow tech companies to monetize their services. The unintended consequence, however, allows targeted ads and malicious posts to become highly effective tools for manipulation [1][17]. Manipulation of attitudes and behavior, in whatever way is desired by the customer (or in whatever ways can be devised by clever misanthropes) is the service. It can cause a user to buy a product, support a cause, or take a more extreme political position. The influence is powerful enough, and insidious enough that it may well have installed a president of a foreign power's choosing in 2016 [1]. This particular unintended consequence is just one example, of course. The tech landscape today is littered with examples where unintended consequences are dramatically more important than those that were intended.

## What is our responsibility?

The predominant view in Silicon Valley, to the extent that harms are acknowledged at all, seems to be that the tech companies whose products gave rise to these destructive impacts should be trusted to address them. In one formulation, the ill-effects of technology can be addressed by the logics of "technological 'solutionism,' meritocracy, and market fundamentalism" [12] In other words, technical solutions to these problems are possible, and will be developed by the best technical talent; market competition will ensure that these solutions are broadly deployed. In sum, the view is that if the market wants ethics, the tech companies will produce "ethicality," much like it would produce any other feature for which there is demand.

Rather than countering technology-induced problems with novel technological solutions, another school of thought advocates integrating ethics into product development by influencing the behavior and thinking of developers [7] and introducing new practices in development organizations [10]. This approach recognizes that ethics is unlike product features, and requires special training and processes to ensure it is folded into product development.

A third perspective argues that we do not have sufficiently developed and globally shared ethical frameworks to address technological harms as ethical decisions [6]. Rather, the issues should be recognized and addressed as fundamentally political questions. By political, Green means "not simply debates about specific political parties and candidates but more broadly the collective social processes that influence rights, status, and resources across society" [6]. Technical decisions with social implications must be subject to deliberation and debate, especially attentive to the voice of those who will be most affected. This view is closely allied to multistakeholder analysis and participatory design.

Regardless of the perspective one takes – whether one of these or views of one's own invention – I take it as a very positive sign that unintended consequences are receiving so much attention. I argue that members of the software engineering research community have particular roles they must play in order to facilitate better outcomes.

Practitioners have courageously taken up the fight against technologies, or applications of technologies, that they find ethically repellant, even at the risk of their jobs. Workers at Amazon[3] and Google[4], for example, have actively protested policies they consider unethical or unfair, even addressing broader societal issues such as climate change[5]. Managers and executives, it seems, are generally too immersed in the for-profit culture to advocate any course of action that would impact the bottom line. Tech workers, on the other hand, are a scarce and valuable asset, and as such wield power. For the moment, it appears that shaping company behavior in an ethical direction rests largely on their shoulders.

Academics, by the nature of our profession, are intimately familiar with the workings of social epistemology. As both as reviewers and as authors who willingly subject their work to the acid bath of the review process, we understand that high standards rigorously enforced are essential for sorting real intellectual progress from self-promotion and wishful thinking. I think the challenge here is staying focused on our real reason for being, i.e., the pursuit of truth – advances backed by credible evidence. We should not only vigorously defend the integrity of the review process for papers and proposals, we should speak out against the uncritical acceptance of unsupported claims circulating online. Perhaps most importantly, we should educate our students, from the very beginning, not just about the substance of our field, but also about how we gather evidence and interpret it, how peer review works and the critical role it plays, and how all forms of knowledge depend on some version of social epistemology.

Finally, I call on us all to begin to think about the ethical implications of what we build, and about ways to do a better job of anticipating potentially harmful unintended consequences of the software we build and deploy. We need to work out ways to measure and observe, to see if these effects are materializing, and plan potential mitigations in advance. One could think of it as a sort of societal scale risk management. I don't think we know much yet about how to do this effectively, but it is overwhelmingly important to figure this out quickly. For the right students, this could make a great dissertation topic.

---

[3] https://gizmodo.com/amazon-workers-demand-jeff-bezos-cancel-face-recognitio-1827037509
[4] https://www.vox.com/recode/2019/11/22/20978537/google-workers-suspension-employee-activists-protest
[5] https://www.washingtonpost.com/technology/2019/09/20/amazon-google-other-tech-employees-protest-support-climate-action/